\documentclass[apjpt4]{aastex}
\shorttitle{Discovery of Blue Hook Stars in M54}
\shortauthors{Rosenberg, Recio-Blanco, \& Garc\'\i a-Mar\'\i n.}

\begin{document}

\title{Discovery of Blue Hook Stars in the Massive Globular Cluster M54}

\author{Alfred Rosenberg}
\affil{Instituto de Astrof\'\i sica de Canarias, Via L\'actea s/n, 
E-38205 La Laguna, Tenerife, Spain}
\email{alf@ll.iac.es}

\author{Alejandra Recio-Blanco}
\affil{Dip. di Astronomia, Universit\`a di
Padova, Vicolo dell'Osservatorio 2, I-35122 Padova, Italy}
\email{recio@pd.astro.it}

\and

\author{Macarena Garc\'\i a-Mar\'\i n}
\affil{Dpto. Astrof\'\i sica Molecular e Infrarroja, Instituto de 
Estructura de la Materia (DAMIR/IEM-CSIC), Serrano 113bis, E-28006
Madrid, Spain}
\email{maca@damir.iem.csic.es}

\begin{abstract}
We present $BV$ photometry centered on the globular cluster M54
(NGC~6715). The color--magnitude diagram clearly shows a blue
horizontal branch extending anomalously beyond the zero age horizontal
branch theoretical models. These kinds of horizontal branch stars
(also called ``blue hook'' stars), which go beyond the lower limit of
the envelope mass of canonical horizontal branch hot stars, have so
far been known to exist in only a few globular clusters: NGC~2808,
$\omega$~Centauri (NGC~5139), NGC~6273, and NGC~6388. Those clusters,
like M54, are among the most luminous in our Galaxy, indicating a
possible correlation between the existence of these types of
horizontal branch stars and the total mass of the cluster. A gap in
the observed horizontal branch of M54 around $T_{\rm eff} = 27\,000$~K could
be interpreted within the late helium flash theoretical scenario, a
possible explanation for the origin of those stars.

\end{abstract}


\section{Introduction}

The horizontal branch (HB) hosts stars with a helium-burning core of
about 0.5 $M_{\odot}$, and a hydrogen-burning shell. The masses of the
hydrogen envelopes vary from more than 0.2 $M_{\odot}$ to less than
0.02 $M_{\odot}$. Furthermore, the less massive the hydrogen envelope
is, the hotter is the corresponding HB star. In the case of a star
cluster, we find a color spread of the HB stars which is called the HB
morphology. To a first approximation, the different color extensions
of observed cluster HBs are described in terms of the variation of
metal abundance, the {\it first parameter} (metal-rich clusters tend
to have short red HBs, while metal-poor ones exhibit predominantly
blue HBs). However, some other parameter (or set of parameters) has
also to be at work, as clusters with nearly identical metallicities
can show very different HB color distributions
\citep{vandenbergh67,sandage67}, leading to the so called
{\it second parameter} debate.

Horizontal branch stars with very low envelope masses ($\leq$ 0.02
$M_{\odot}$, $T_{\rm eff} > 20\,000$~K), known as extended or ``extreme HB''
(EHB) stars, are probably the most extreme expression of the second
parameter problem. They have lost up to twice the mass during the red
giant branch (RGB) ascent than other HB stars in the same cluster
\citep{dcruz96}. As a result, in contrast to the more massive blue HB
stars, EHB stars do not ascend the asymptotic giant branch (AGB) but
evolve directly onto the white dwarf domain \citep{sweigart74}.
Recently, \citet{whitney98} and \citet{dcruz00} revealed the existence
of a particular kind of EHB star: a population of hot subluminous HB
stars, lying up to 0.7 mag below the ZAHB and forming a hook-like
feature in the far-UV color--magnitude diagram (CMD) of
$\omega$~Centauri. These ``blue hook'' stars have effective
temperatures up to $40\,000$~K and cannot be produced by canonical HB
evolution \citep{brown01}. In the optical, for effective temperatures
higher than 10\,000~K, ultraviolet radiation constitutes the main part
of the energy flux coming from the stellar surface, making the HB, in
practice, vertical in the classical $V$ {\it vs.} ($B-V$) plane become
of bolometric correction. Hence, in optical CMDs blue hook stars are
located at the faintest extreme of the HB.

In this letter, we present $BV$ photometry centered on the globular
cluster (GC) M54. The CMD clearly shows a blue HB anomalously
extending beyond ZAHB models. Previous photometric studies of this
cluster, in ($V-I$), were not suitable for properly revealing this
extremely hot stellar population. Initially, blue hook stars were
detected in the clusters NGC~2808 and $\omega$~Centauri. More
recently, \citet{busso03} have reported their presence also in the
blue HB tail of NGC~6388. In addition, as noted by \citet{brown01},
the CMD of NGC~6273 shown by \citet{piotto99} shows a blue HB
extending to $M_V$ $>$ 5, and therefore beyond theoretical ZAHB
models. All these clusters are among the most massive GCs in the
Galaxy, as well as M54, which is the second most massive GC known. In
Section 2, we describe the observations and the photometric reduction
techniques. In Section 3, we analyze the extended HB of M54 and its
interpretation inside the late helium flashers scenario. Finally, in
Section 4, we summarize the results and consider their wider
implications.

\section{Observations and Reduction}

The observational data base consist of four images, two in $B$ and two
in $V$, of 30 and 900 s each, centered on M54. Images were observed in
service mode on 2002 July $8$ at the ESO 3.5 m New Technology
Telescope (NTT), with a seeing of $\sim0.5$ arcsec. The detector, the
Superb Seeing Imager (SUSI-2), is a mosaic of two 2k $\times$ 4k EEV
CCDs, with a size of $0.08$ arcsec per pixel and a total field of view
of $5.5$ $\times$ $5.5$ arcmin that were binned $2\times2$.

The images were corrected for bias and spatial sensitivity variations
using the respective master flats, computed as the median of all
available sky flats of the specific run. Afterward, photometry was
performed using the DAOPHOT\-/\-ALLSTAR\-/\-ALLFRAME software
\citep{stetson87,stetson94}.

The absolute calibration of the observations, which include the $BVRI$
filters, will be published in a forthcoming paper. It is based on six
fields of standard stars from the catalog of \citep{landolt92}, and
the absolute zero-point uncertainties of our calibration are
$\leq0.02$ mag for each of the four bands.

\section{The extended HB of M54 and its interpretation}

Figure 1 shows the CMD of M54 in an area away from the crowded region
with $R$ $<$ 90 arcsec, where $R$ is the projected distance from the
cluster center. We can identify at least three stellar systems: the GC
M54, the Sagittarius Dwarf Galaxy, and the Milky Way bulge
\citep{layden00}. All stars plotted were selected using the sharp
parameter ($-$0.25 $<$ sh $<$ 0.25) and the $V$ error ($\leq0.1$). In
this paper we are interested in M54, for which the main sequence (MS),
RGB, and HB are clearly visible. The whole CMD has been shifted in
color and magnitude by $E(B-V)=0.16$ and $(m-M)_0=17.25$ in order to
fit the ZAHB model. The CMD shows an extended HB, which spans almost
4.75 mag in $V$ and extends down to $M_V\simeq$ 5.25, i.e., $\sim$ 1.5
mag below the turn-off. The thick line shows the ZAHB model by
\citet{cassisi97} for a metallicity of [Fe/H] $=$ $-1.31$.

Crowding and completeness experiments where also performed along the
HB sequence. The completeness was found to be higher than $50\%$ for
the entire CMD shown ($60\%$ and $70\%$ zones are labeled in Fig. 1.)

From the crowding experiments, we have obtained the photometric
dispersion for synthetic stars located on the HB theoretical model
line, to which we have add three times the typical dispersion in color
(0.025~mag) measured from the region between $1.0\leq{M_V}\leq4.5$
of the HB. These limits are represented by the dashed lines and all 78
stars within these two lines are considered as the HB stars in the
following discussion.

Although, because of bolometric correction, the ($B-V$) color is not a
good temperature indicator for the hot HB stars, the CMD clearly shows
a horizontal branch $\sim$ 0.5 mag deeper than the ZAHB model. At
least nine stars in Fig. 1 seem to have effective temperatures higher
than $33\,600$~K, which is the end of the ZAHB model by
\citet{cassisi97}. As mentioned in the introduction, a different
evolutionary path for populating the very hot end of the HB is
therefore needed. One possible explanation is that these stars
experience a delayed helium core flash. \citet{castellani93} and
\citet{dcruz96,dcruz00}, showed that even if a star suffers
a very high mass loss during its RGB phase, it can still ignite He
burning after evolving off the RGB.

If the helium flash occurs somewhere between the tip of the RGB and
the top of the helium dwarf cooling curve ({\it early} hot flashers),
the star will settle onto the EHB without any mixing between its
helium core and hydrogen envelope, following a canonical evolutionary
path to the EHB. The reason for this can be found in the entropy
barrier caused by the hydrogen-burning shell, which prevents the
convection zone produced by the helium flash from penetrating into the
hydrogen envelope \citep{iben76}. Early hot flasher models by
\citet{brown01} predict that these stars reach a maximum temperature
of $31\,500$~K on the EHB, and that their luminosities are almost
indistinguishable from the luminosities of canonical EHB
stars. \citet{brown01} claimed that if the helium flash occurs while
the star is descending the white dwarf cooling curve ({\it late} hot
flashers), the entropy barrier carried by the hydrogen-burning shell
becomes a too weak to prevent mixing between the core and the star's
envelope. The resulting star will have a temperature of about
$37\,000$~K and will lie as much as 0.7 mag below the ZAHB in the {\it
UV} CMD, further supporting the argument that blue hook stars are the
progeny of late hot flashers. The fact that blue hook stars are both
hotter and more helium-rich than classical EHB stars has been
observationally confirmed by the spectroscopic analysis of
\citet{moehler02} in $\omega$~Centauri. In addition, full evolutionary
computation of helium flash induced-mixing in Population II stars has
recently been developed by \citet{cassisi03}. They modeled the
incursion of the He flash-driven convective zone into the H-rich
external layers, with a subsequent surface enrichment in He and C. In
agreement with the observations, models experiencing this dredge-up
event are significantly hotter than their counterparts with H-rich
envelopes. They also compare their abundance predictions of He with
measurements by \citet{moehler02}.
 
On the other hand, the sharp transition between the early and the hot
flashers, corresponding to a difference in mass loss of only 10$^{-4}$
$M_{\odot}$ along the RGB, would produce a gap in the observed stellar
distribution. The CMDs of $\omega$~Centauri by \citet{kaluzny97} and
\citet{lee99}, and NGC~6273 by \citet{piotto99} do not seem to show an
EHB gap. \citet{brown01} suggest that the gap in $\omega$~Centauri is
perhaps blurred by the metallicity distribution of the cluster. The
differential reddening and magnitude limit could be the reason in the
case of NGC~6273. They test their blue hook models by trying to
reproduce the luminosity function of NGC~2808 by \citet{bedin00}. They
first assume that the EHB and the blue hook zone are uniformly
populated and then include the HB evolution of the models to brighter
$V$ magnitudes. They find that the gap in the observed ZAHB
distribution corresponds to the gap between the canonical ZAHB and
their blue hook models with mixed envelopes, further supporting the
scenario of the late hot flashers.

In Figure 2, we compare the stellar distribution in $M_V$ of the M54
horizontal branch from our photometry (shown as filled triangles) with
that of NGC~2808 by \citet{bedin00} (open circles). The total number
of stars in the NGC~2808 distribution (160) was normalized to that of
M54 (78) for a better comparison. For the latter cluster, a distance
modulus of ($M-m$)$_V=$ 15.74 \citep{bedin00} was used. Error bars for
M54 correspond to the square root of the number of the stars per bin,
and the bin size is 0.30 mag. Interestingly, M54 presents a gap in the
stellar distribution at the same absolute $V$ magnitude ($M_V$ $\sim$
4.3) as NGC~2808 (marked in Figure 2 with a vertical arrow). As
pointed out before, in the late helium flasher scenario, this gap
would separate the canonical EHB from the blue hook stars and would be
only apparent because of the mixed envelopes of blue hook stars
\citep{brown01}. If this is true, a quite abundant blue hook
population is clearly seen in M54 at the same absolute magnitudes
(effective temperatures) as in NGC~2808. From this histogram, we found
that the number of star in the blue hook region, compared to the whole
HB star sample (for $1.0 \leq M_V \leq 5.5$) is $35\%$ in the case of
M54 and less than $20\%$ in the case of NGC~2808. We do not know if
this difference is caused by the metallicities, which are very similar
for both clusters ([Fe/H] $= -1.25$ and [Fe/H] $= -1.11$ for M54 and
NGC~2808, respectively, in the Carretta \& Gratton 1997 scale
\citep{rutledge97}), which helps the direct comparison of their
HBs. On the other hand, it would again be necessary to take into
account the HB evolution of the models to brighter V luminosities in
order to explain the location of the gap at $T_{\rm eff}\sim 27\,000$~K, as
already claimed by \citet{brown01} for the case of NGC~2808. On the
other hand, we note that the first gap in the HB of NGC~2808 pointed
out by \citet{bedin00} at $V\sim18.5$ mag ($M_V\sim2.8$ mag) is not
visible in M54. On the other hand, there is an underpopulated zone at
lower magnitudes around $M_V\sim3.2$ mag.

Finally, the peculiar bimodality of the NGC~2808 HB morphology, which
includes both a blue HB and a red HB clump, may also be present in
M54. This would increase the similarities in the peculiar features
observed in the HB of these two clusters. However, although there is
evidence for a red HB both in our diagram and already published $VI$
photometry (see, for example, \citet{sarajedini95}), this feature
could belong to the Sagittarius (Sgr) dwarf galaxy field. 
\citet{layden00}, in their extensive $VI$ photometry of M54 and
the Sgr galaxy, conclude that their subtractions of the Sgr field from
the M54 CMD cannot ascertain beyond doubt whether the anonymous red HB
belongs to Sgr or to M54. 

On the other hand, in the recent work by \citet{monaco03}, it is clear
that the Sgr dwarf galaxy also possesses an old and relatively
metal-poor stellar component that populates the blue HB
tail. Interestingly, \citet{monaco03} find that the the population of
faintest ($V>18.6$) blue HB stars is relatively lower in the Sgr field
than in M54. Moreover, the spread of the stars in the hottest part of
the Sgr field HB is rather high.  In our Figure 1 we can see that the
stars that we present as blue hook stars are clearly grouped in a
relatively small region of the CMD.  They are clearly separate from
the MS, the rest of our CMD being extremely clean. For all these
reasons we believe that the stars in our CMD, located under the GAP,
are really cluster members.

\section{Discussion}

As described in the previous section, we can conclude that the M54
horizontal branch hosts a blue hook stellar population that extends
the HB to fainter magnitudes than ZAHB models. Understanding the
origin of hot HB stars is important not only for our fine-tuning of
stellar evolution theory, but has wider applications in
astrophysics. Indeed, hot HB are now considered to be the prime
contributors to the ultraviolet emission in elliptical galaxies
\citep{greggio90,brown01}. Blue hook stars are not, however, a common
feature of all GCs with an extended blue HB morphology (see for
instance \citep{moehler00}, for the cluster NGC~6752).

At this point, it might be useful to examine the similarities in the
physical properties of this cluster with the other GCs with already
confirmed or suspected blue hook stars in their HBs---that 
is NGC~2808, NGC~6388, NGC~6273, and $\omega$~Centauri---in order to shed
more light on the origin of this peculiar kind of star. The most
striking analogy between these five clusters is that they are among
the most massive clusters in our Galaxy. Such characteristics would
explain the presence of a larger EHB population, but would not, in
principle, be directly considered as a justification for the bluer HB
morphology of these clusters. The total absolute $V$ magnitude ($M_V$)
can give a good estimate of these clusters' total luminosities (as all
GCs have similar color indices and hence similar bolometric
corrections) and therefore a good measurement of the baryonic mass of
these old stellar systems. From (\citet{harris96}, updated to the new
catalog version of 2003 February) we find $M_V$ $=$ $-9.18$ for
NGC~6273, $-9.39$ for NGC~2808, $-9.42$ for NGC~6388, $-10.01$ for M54,
and $-10.29$ for $\omega$~Centauri.

Nevertheless, we could think that if there is a distribution in mass
loss along the RGB, then the high mass loss tail of this distribution
would be more likely to be occupied in a more massive cluster. If this
is true, a correlation between the number of hot HB stars per stellar
mass in a cluster and its HB extension and/or total mass at constant
metallicity should also exist. The first results of a multivariate
analysis (Recio-Blanco et al., in preparation), based on a photometric
database of 74 GCs \citep{piotto02} seem to exclude such correlation.

On the other hand, both M54 and $\omega$~Centauri are suspected of
being the nuclei of a current and former dwarf galaxy,
respectively. In fact, M54 must play an important role in the star
formation history of the Sagittarius galaxy, as it lies in the high
density region of Sgr \citep{ibata97}, and, as pointed out by
\citet{layden00}, it marks one of the earliest epochs of star
formation in Sgr. M54 may be similar to the nuclear star clusters seen
in nucleated dwarf elliptical galaxies \citep{sarajedini95}. On the
other hand, because of the unusual properties of $\omega$~Centauri
(mainly abundance variations and metallicity spread), the scenario
that this cluster may be also the core of a disrupted dwarf galaxy
(e.g., \citet{freeman93}) had a continuous infall of gas to its
center, leading to a variable star formation history, is becoming
popular. Moreover, the NGC$~$2808 HB bimodality could be interpreted
within a similar scenario of cluster self enrichment if we consider
the \citet{dantona02} suggestion of the influence of a possible second
generation of He-rich stars in the final cluster HB morphology. In
this sense, the correlation between the high mass of these clusters
and the existence of blue hook stars (and so of their progenitors as
the proposed late hot helium flashers) could also be linked to the
second parameter debate regarding the more general problem of GC HB
morphology.

It is interesting to point out, in addition, that age differences from
cluster to cluster would not be enough to explain the second parameter
problem. This is the case for NGC$~$2808, coeval with other clusters
of similar metallicity but much shorter HBs such as NGC~362, NGC~1261,
or NGC~1851, all of them still $20\%$ younger than NGC~288, NGC~5904,
or NGC~6218 \citep{rosenberg99,rosenberg00a,rosenberg00b}. In fact,
many other massive GCs of different metallicities show particularly
extended HB morphologies: NGC$~$6266 ($M_V = -9.19$, [Fe/H] $\sim
-1.3$), NGC$~$2419 ($M_V = -9.58$, [Fe/H] $\sim -2.1$=) and NGC$~$6441
($M_V = -9.64$, [Fe/H] $\sim -0.5$). Deeper photometry in the blue or
in the ultraviolet could reveal the presence of a blue hook population
for those clusters as the one already detected in NGC$~$2808,
$\omega$~Centauri, NGC~6388, NGC~6273, and now in M54.

\acknowledgments

We would like to thank our anonymous referee for many valuable
comments and suggestions. This report is based on observations with
the ESO {NTT + SUSI2}, located at the La Silla Observatory, Chile (ESO
proposal 69.D-0655(A)). We thank L. R.\ (``Rolly'') Bedin for
providing the NGC~2808 HB photometry and Santino Cassisi for his ZAHB
models. We are grateful to Antonio Aparicio, Carme Gallart, Giampaolo
Piotto and Ivo Saviane for allowing us to used these data, from a
larger project entitled ``Relative Ages of Outer Halo Globular
Clusters'', in advance of publication. ARB recognizes the sustenance
from {MIUR} and from {ASI}.

\clearpage

\begin{figure}
\epsscale{0.7}
\plotone{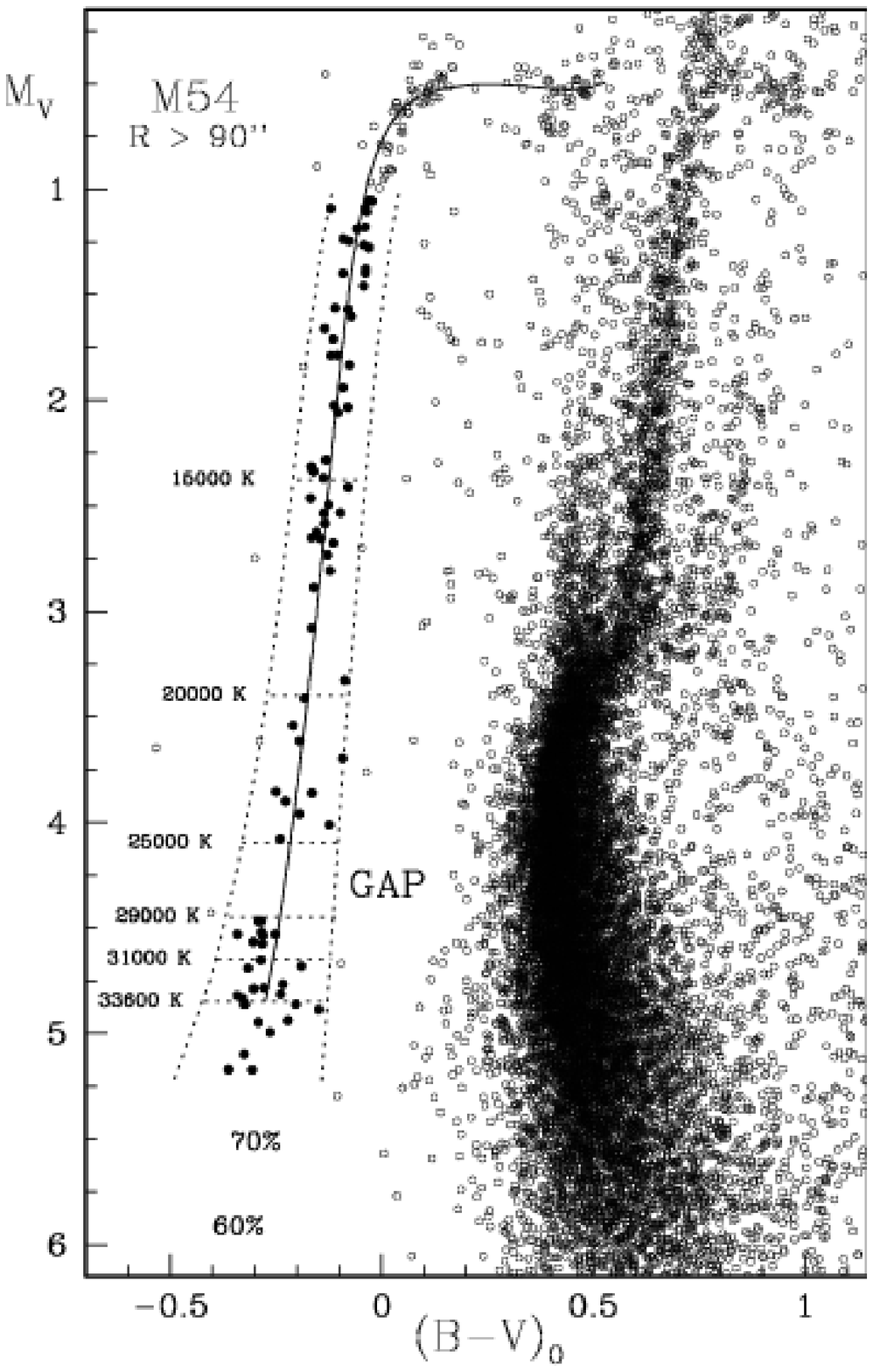}
\caption{The $M_V$ {\it vs.} $(B-V)_0$ CMD for 15\,202 stars from the 
observed field centered on M54. All stars plotted were selected by the
sharp parameter ($-$0.25 $<$ sh $<$ 0.25) and the $V$ error ($\leq$ 0.1),
and are all farther than 90 arcsec from the cluster center. The
overplotted thick line is the ZAHB model by Cassisi \& Salaris (1997),
for a metallicity of [Fe/H] $= -1.31$. On both sides, as dashed lines,
are shown the color limits of the region within which we have adopted
stars as HB stars. Several temperatures, obtained from the ZAHB model,
are also shown. Note the GAP between 25\,000~K and 29\,000~K. The 27 stars
below the GAP and between the dashed lines are expected to be the blue
hook stars. The 70\% and 60\% completeness levels, obtained from a
crowding experiment along the HB region, are also
labeled. \label{fig1}}
\end{figure}

\clearpage

\begin{figure}
\epsscale{0.8}
\plotone{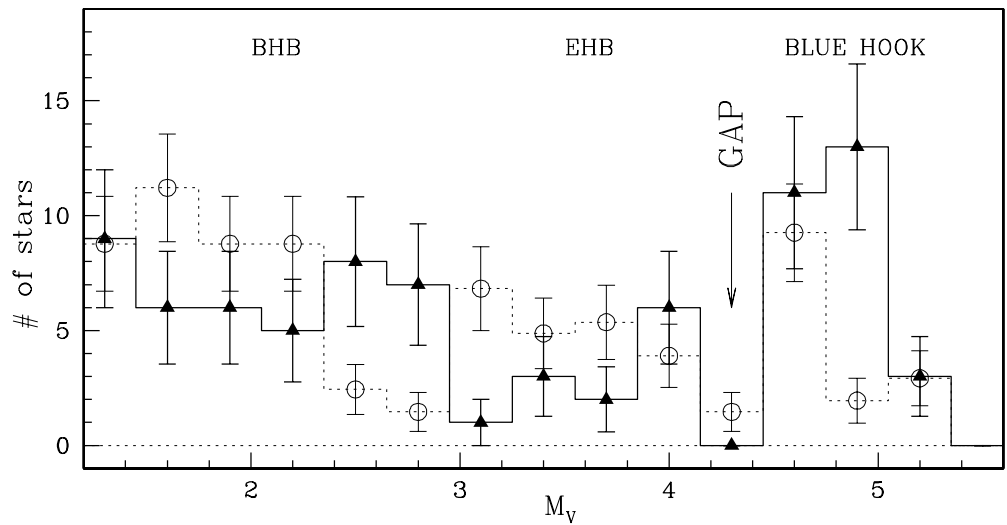}
\caption{Stellar distribution of the M54 HB (filled triangles) 
from our photometry and NGC~2808 (open circles) from Bedin et
al. (2000). The same gap at $M_V$ $\sim$ 4.3 is observed, possibly
separating the canonical HB from blue hook stars.
\label{fig2}}
\end{figure}

\end{document}